# Ch. XX – Ultrafast X-Ray Scattering: New Views of Chemical Reaction Dynamics

Peter M. Weber[1], Brian Stankus[2] and Adam Kirrander[3]

**Abstract** The advent of ultrafast pulsed X-ray free-electron lasers with very high brightness has enabled the determination of transient molecular structures of small and medium-sized organic molecules in excited states and undergoing chemical dynamics using X-ray scattering. This chapter provides an introduction into X-ray scattering theory and considers several important aspects relating to the experimental implementation. Ultrafast gas phase X-ray scattering is shown to provide new observables to elucidate the dynamics of chemical reactions by providing complete, time-dependent molecular structures. Consideration of correlations between structural parameters are important for molecules far from their equilibrium and the changes in electron density distributions of molecules upon optical excitation need to be considered in the analysis. Future technological developments are expected to lead to further important advances.

## XX.1 Introduction

The determination of molecular structures is a foundational achievement of 20th century chemistry. Started as a curiosity-driven inquiry in chemistry, molecular structures now form the bedrock of many fields including molecular biology, pharmacology and material science. Applications in medicine, drug design, information storage, communication devices and many other areas have immeasurable benefit for modern society. Instrumental advances at the beginning of the 21st century have now made it possible to determine molecular structures in excited states. Coupled with a time resolution deep into the femtosecond regime, we can now measure time-evolving molecular structures of chemical systems during vibrational motions or chemical reactions. The measurement of molecular structures far from equilibrium will advance many fields, possibly including the selective control of chemical reactions, which might lead to the synthesis of yet unknown materials with useful molecular properties and/or the prevention of undesirable reactions. Just as knowledge of ground state molecular structures has been invaluable to the development of computational methods that now find widespread application in the molecular sciences, data on excited state and transient molecular structures can serve as benchmarks for the refinement of quantum chemical methods.

[1] *Department of Chemistry, Brown University, Providence, Rhode Island 02912, USA*

[2] *Department of Chemistry, Western Connecticut State University, Danbury, Connecticut 06810, USA*

[3] EaStCHEM *School of Chemistry and Centre for Science at Extreme Conditions, University of Edinburgh, David Brewster Road, Edinburgh EH9 3FJ, United Kingdom*

For the determination of static molecular structures in classical chemistry, X-ray diffraction is an essential tool. X-ray and electron scattering patterns are Fourier transform projections of the molecular electron and charge density distributions, respectively. Both X-ray and electron scattering are sensitive to the nuclear coordinates and the electron density distributions. Electron scattering results from electrostatic interaction of the electrons in the electron beam with the nuclei and the electrons of the molecule. X-ray scattering arises from the interaction of a molecule's electrons with the electromagnetic field, and therefore only depends on electron density distributions. Since the atomic core electrons are tightly centered on the nuclei, X-ray scattering also determines nuclear geometries. Thus, both scattering experiments can yield the molecular geometry, as well as the associated electron density distributions. Unlike in time resolved spectroscopies, where the Heisenberg uncertainty relation $\Delta E \cdot \Delta t \geq \hbar/2$ imposes a fundamental limitation, scattering experiments measure spatial structures that can be measured with temporal resolution without being restricted by an uncertainty relation. The prospect of measuring nuclear geometries and electron density distributions, *i.e.* chemical bonding, with ultrafast time resolution makes time-resolved scattering an attractive field to develop. The high sensitivity of scattering experiments makes it possible to measure molecular systems even in dilute vapors.

## XX.2 Elements of Scattering Theory

### XX.2.1 X-ray scattering

Scattering of hard X-rays can be treated in the perturbative regime[1] using the first Born and the Waller Hartree approximations[1,2]. The detectors used in ultrafast scattering are generally not energy-resolved, which means that we are mainly concerned with total scattering. The differential cross section for total scattering is[3],

$$\frac{d\sigma}{d\Omega} = r_0 |\boldsymbol{e}_0 \cdot \boldsymbol{e}_1|^2 S(\boldsymbol{q}), \qquad (XX.1)$$

where $r_0 = e^2/m_e c^2$ is the so-called classical electron radius ($e$ signifies the charge and $m_e$ the mass of an electron, $c$ the speed of light). The polarization factor $|\boldsymbol{e}_0 \cdot \boldsymbol{e}_1|^2$ accounts for the polarization of the incoming X-rays via,

$$|\boldsymbol{e}_0 \cdot \boldsymbol{e}_1|^2 = \begin{cases} 1 & \text{vertical} \\ \cos 2\theta & \text{horizontal}, \\ \dfrac{(1 + \cos^2 2\theta)}{2} & \text{unpolarized} \end{cases} \qquad (XX.2)$$

for unpolarized sources or sources polarized in the vertical or horizontal scattering planes. The dynamic structure factor is given by

$$S(\boldsymbol{q}) = \sum_{\beta} |\langle \psi_\beta | \hat{L} | \psi_\alpha \rangle|^2 = \langle \psi_\alpha | \hat{L}^\dagger \hat{L} | \psi_\alpha \rangle, \qquad (XX.3)$$

where $|\psi_\beta\rangle$ and $|\psi_\alpha\rangle$ are the final and initial electronic states and the sum includes all states. The second equality constitutes a powerful result that makes it possible to calculate the total scattering for a particular electronic state (in this instance $|\psi_\alpha\rangle$) without reference to any other states[3]. The scattering vector $\boldsymbol{q}$ (a. k. a. the momentum transfer vector) is given by the difference between the incident and the scattered wave vectors,

$$\boldsymbol{q} = \boldsymbol{k_0} - \boldsymbol{k_1}, \qquad (XX.4)$$

where $k_0 = |\boldsymbol{k_0}| = \frac{2\pi}{\lambda} = \frac{2\pi E}{hc}$ with $\lambda$ and $E$ the wavelength and energy, respectively, of the incoming X-rays. The scattering angle $2\theta$ is related to the magnitude of the scattering vector via $q = |\boldsymbol{q}| = |\boldsymbol{k_0} - \boldsymbol{k_1}| = 2k_0 \sin\theta$. Finally, the scattering operator in Eq. XX.3 is given by,

$$\hat{L} = \sum_{j=1}^{N_e} e^{i\boldsymbol{q}\boldsymbol{r}_j}, \qquad (XX.5)$$

where the sum runs over all the $N_e$ electrons with coordinates $\boldsymbol{r}_j$. One may decompose the contributions to the dynamic structure factor into elastic and inelastic components,

$$S(\boldsymbol{q}) = |F(\boldsymbol{q})|^2 + S_{\text{inel}}(\boldsymbol{q}), \qquad (XX.6)$$

where the elastic component $|F(\boldsymbol{q})|^2$ is proportional to the form factor,

$$F(\boldsymbol{q}) = \langle \psi_\alpha | \hat{L} | \psi_\alpha \rangle = \int \rho_\alpha^{(N_e)}(\boldsymbol{r}) e^{i\boldsymbol{q}\boldsymbol{r}} d\boldsymbol{r} \qquad (XX.7)$$

with $\rho_\alpha^{(N_e)}(\boldsymbol{r})$ the electron density. The inelastic component, $S_{\text{inel}}(\boldsymbol{q})$, is what remains to make up the total scattering and accounts for transitions to all states $\alpha \neq \beta$. Quite often, rotationally averaged signals are considered[4,5], especially in the gas phase, in which case only the amplitude $q$ matters.

## XX.2.2 Independent Atom Model

The independent atom model (IAM), originally proposed by Debye[6], is widely used. It approximates the electron density as a sum of isotropic isolated-atom electron densities centered at the positions of the nuclei[3,7], which makes it possible to use conveniently tabulated[8] atomic form factors, $f_i(q)$, and inelastic corrections, $S_{\text{inel},i}^{IAM}(q)$, to express the rotationally averaged total scattering as,

$$S_{IAM}(q) = \sum_{i=1}^{N_{at}} \sum_{j=1}^{N_{at}} f_i(q) f_j(q) \frac{\sin(qR_{ij})}{qR_{ij}} + \sum_{i=1}^{N_{at}} S_{inel,i}^{IAM}(q) \qquad (XX.8)$$

where the indices $i$ and $j$ run over the $N_{at}$ atoms with $R_{ij}$ the distance between pairs of atoms. As previously in Eq. XX.6, the first term is the elastic component, while the second is the inelastic component. For IAM, the inelastic component is considered independent of the molecular geometry. The shortcomings of the IAM approximation are well-documented[9,10,11,3,5], but it remains a reasonable approximation for many molecular systems, especially in the electronic ground state, and constitutes an effective way to model the most prominent features of molecular scattering patterns.

### XX.2.3 Comparison to electron scattering

Most molecular structures known today were determined with either X-ray or electron diffraction. Electron diffraction is closely related to X-ray scattering, but arises from the Coulomb interactions between charged particles, meaning that the electrons scatter from both target electrons and nuclei[12,13,14]. In practice, the Thomson cross section above is replaced by the Rutherford cross section and an additional $q^{-4}$ damping appears in the resulting scattering intensity[15,16,17,18]. Electron form factors can be computed from the Fourier transform of the atomic potentials[15,8,19,20] or found as approximate analytical expressions[21,22]. It is most convenient to use the Mott-Bethe formula, which relates electron scattering form factors to X-ray atomic scattering factors,

$$f_e(q) = \frac{2m_e e^2}{\hbar^2} \cdot \frac{Z_i - f_i(q)}{q^2} \qquad (XX.9)$$

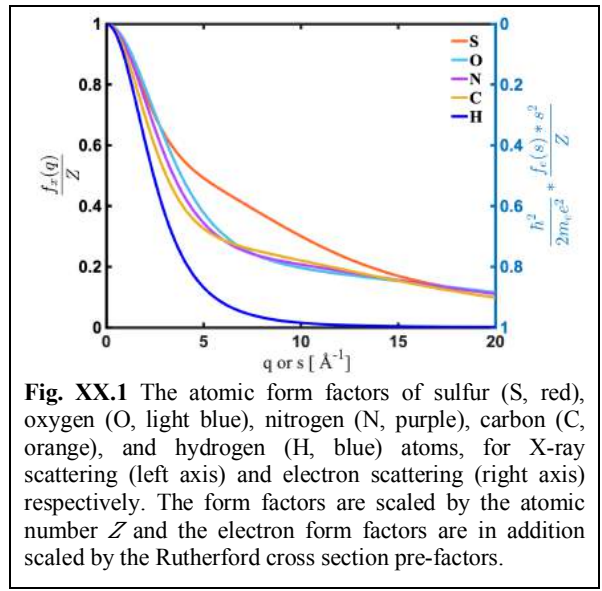

**Fig. XX.1** The atomic form factors of sulfur (S, red), oxygen (O, light blue), nitrogen (N, purple), carbon (C, orange), and hydrogen (H, blue) atoms, for X-ray scattering (left axis) and electron scattering (right axis) respectively. The form factors are scaled by the atomic number $Z$ and the electron form factors are in addition scaled by the Rutherford cross section pre-factors.

with $\hbar$ the reduced Planck's constant and $Z_i$ the atomic number of each atom. Note that we follow the X-ray convention and use the symbol $q$ rather than $s$ for the momentum transfer. Equation XX.9 suggests that outside of the constants and the $\frac{1}{q^2}$ dependence of the Rutherford scattering amplitude, electron and X-ray scattering have similar information content in that both depend on the X-ray form factor. Yet they are also complementary: the X-ray form factors decrease with increasing scattering angle while the electron form factors increase, Fig. XX.1. The increasing electron form factors partially compensate for the loss of electron scattering signal arising from the rapidly decaying $\frac{1}{q^4}$ term of Rutherford scattering.

Several important conclusions follow from this discussion: First, the cross section for electron scattering is dramatically larger than that for X-rays. This is the reason why early studies explored ultrafast electron diffraction (UED) to study excited state structures and dynamics.[23,24,25,26,27,28,29,30] The first UED measurement of the ring-opening reaction of 1,3-cyclohexadiene (CHD) was conducted almost 20 years ago.[31,32] Even so, at the time the experiments were difficult because space-charge interactions between the electrons, which are closely confined within an ultrashort electron pulse, make it challenging to achieve the excellent signal-to-noise ratio demanded by the rapid ($1/q^4$) dependence of the scattering signal on the momentum transfer vector $q$. Tremendous progress results from using relativistic electrons[33,34] and pulse compression techniques.[35,36] New MeV ultrafast electron diffraction instruments have yielded important advances, starting with the dynamics of $I_2$ molecules.[37] More recent experiments include the applicability to larger systems such as CHD.[38] Electron diffraction with ultrafast time resolution continues to be a valuable complement to the X-ray scattering. In a current review article, we delineate the accomplishments and promise of UED.[39]

The larger cross section for electron scattering is valuable, but it only goes so far. The tremendous brightness of modern X-ray free electron lasers (XFEL) beams has now turned the table: while the electron scattering cross section is larger by a factor of $10^5$, there are $3·10^7$ more photons in an X-ray pulse than there are electrons in the electron pulse. Consequently, the scattering signal is about 300 times larger for X-rays. Since this comparison is on a pulse-by-pulse basis, the difference will become even more dramatic as high repetition rate XFEL technology advances.

## XX.3 Experimental Implementation of Pump-Probe X-Ray Scattering

The experimental scheme for the ultrafast gas-phase X-ray scattering experiments is conceptually straightforward. An optical pump laser pulse is focused onto the gaseous target and initiates the photochemistry, Fig. XX.2. It is followed in time by an X-ray probe pulse that produces a scattering image on an area detector. The scattering signal is recorded for variable time delays to monitor the dynamical structure of the molecules. In its implementation, the experiment requires a wide array of considerations affecting the final observed signal. Important design features relevant for the initial study of the ring-opening of 1,3-cyclohexadiene[40] were described in a detailed methods paper in 2016[41]. Since then, the experiment has been improved and implemented at the Coherent X-ray Imaging (CXI) endstation of the Linac Coherent Light Source (LCLS).

Gas-phase X-ray scattering experiments have been enabled by the rapid advancement in capabilities of XFELs such as LCLS. The high photon energies coupled with exceptional brightness (see Sect. XX.2.3) yield high signal-to-noise ratios that allow for recovery of detailed structural information. In addition to the pulse characteristics of LCLS, the CXI endstation offers several other distinct advantages. Firstly, the interaction and detection regions are fully in-vacuum, which helps to eliminate unwanted background scattering signals from air. In addition, CXI houses multiple differentially pumped sample chambers, which allows for additional measurements to be made downstream of the initial CSPAD detector. Both of these features are important for improving the signal-to-noise ratio of the measured scattering patterns. Of course, many other aspects of the experimental design, some of which are discussed below, need careful optimization to ensure stable and reliable signals.

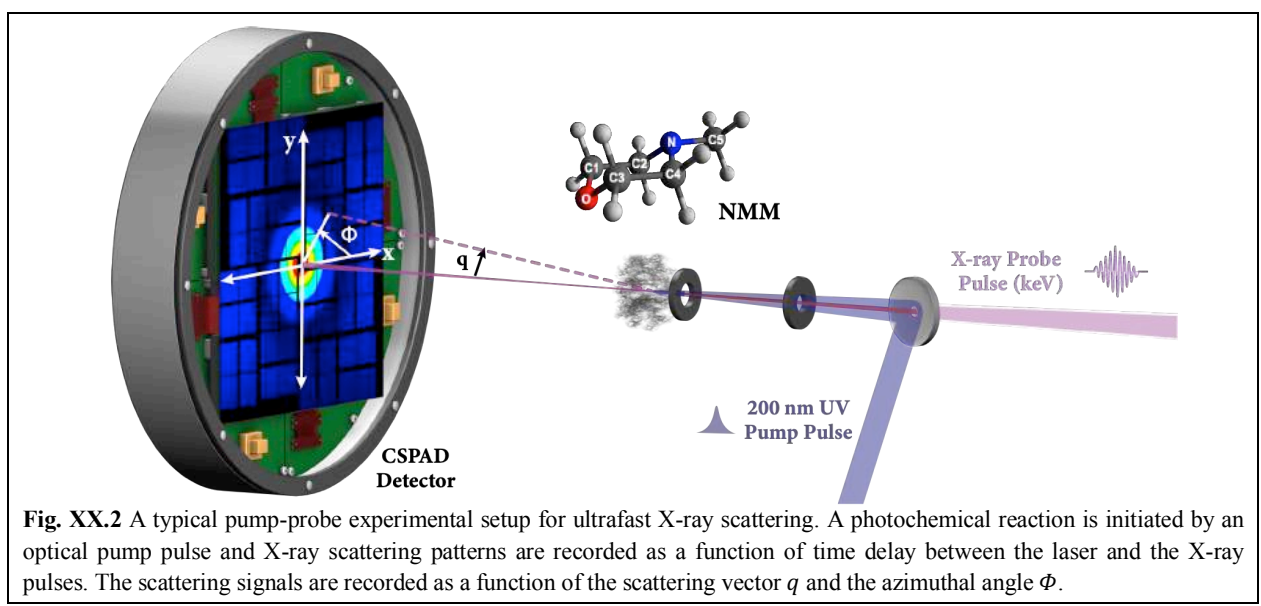

**Fig. XX.2** A typical pump-probe experimental setup for ultrafast X-ray scattering. A photochemical reaction is initiated by an optical pump pulse and X-ray scattering patterns are recorded as a function of time delay between the laser and the X-ray pulses. The scattering signals are recorded as a function of the scattering vector $q$ and the azimuthal angle $\Phi$.

**XX.3.1 Scattering Cell Design**

The scattering cell shown in Fig. 3 is a 2.4 mm pathlength stainless steel gas cell, with a 3.2 mm inner diameter inlet tube for the gaseous target molecule. The cell was constructed with a 250 μm platinum entrance aperture on the upstream side and a beryllium exit window with a 250 μm aperture on the downstream side. This avoids obstruction of the primary X-ray beam and allows for sufficient gas flow between X-ray pulses (see below).

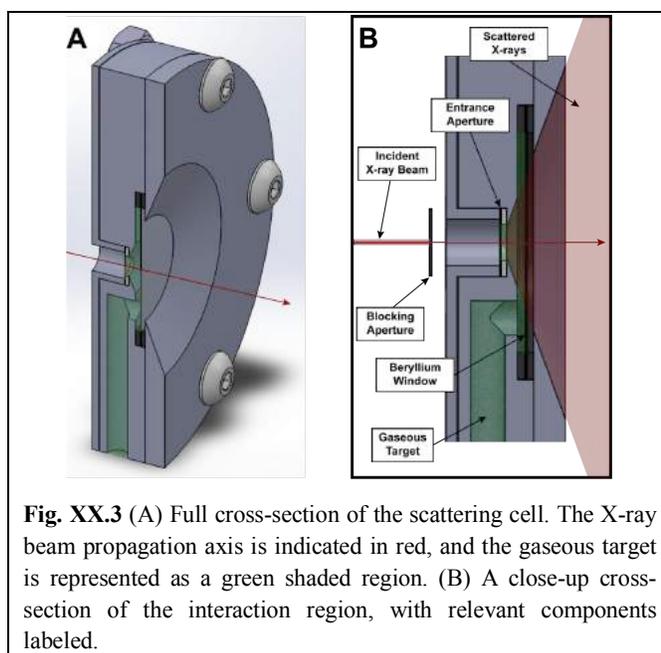

**Fig. XX.3** (A) Full cross-section of the scattering cell. The X-ray beam propagation axis is indicated in red, and the gaseous target is represented as a green shaded region. (B) A close-up cross-section of the interaction region, with relevant components labeled.

In order to prevent Bragg scattering from the primary X-ray beam on the cell entrance aperture, an upstream blocking aperture is used (see Fig. XX.3B). The upstream blocking aperture has a 200 μm diameter opening, which allows the vast majority of the primary X-ray beam (nominally focused to 30 μm FWHM) to pass through while blocking the low-intensity "edges" of the incident beam. This inevitably causes Bragg scattering from the platinum metal, which is subsequently blocked from entering the interaction region by the scattering cell itself. The entrance aperture, which has a 250 μm diameter opening, then allows the primary X-ray beam to pass cleanly to the interaction region. The diffuse scatter from the blocking aperture that enters the cell through the entrance aperture is weak and occurs at such a small scattering angle that it is not detected in the experiment.

The cell was designed to allow scattering at angles up to ~60° to exit the interaction region unobstructed, with the exception of the beryllium exit window. A very thin (100 μm thick) disk of beryllium was chosen as the exit window material because it is nearly transparent to X-rays (98.8% transmission at 9.5 keV). There is a very small dependence of the transmission on the scattering angle due to the changing path length. This effect, which is on the order of 0.5% in the absolute scattering signal, is taken into account during the data analysis.

The relatively small 2.4 mm pathlength of the interaction region was chosen for two reasons. Firstly, the short interaction length controls the Beer-Lambert attenuation of the UV pump pulse as it propagates through the sample. During the pump-probe experiments, it is necessary to have a significant number of excited molecules (as this is the signal being measured) while avoiding an excitation probability of >10% at any point in the interaction region to avoid multiphoton absorption. To attain both of these conditions, a near-constant excitation probability of less than 10% is desired. To achieve this, the Beer-Lambert attenuation is offset by weakly focusing the pump beam at the downstream end of the cell as detailed in a previous report[41]. The short interaction length, in concert with careful control of the gas pressure and UV intensity, helps to ensure that the desired balance is achieved.

The other benefit of a short interaction length is an improved resolution of the scattering angle. With any finite interaction length, there is an inherent limit on scattering angle resolution caused by scattering from molecules at the upstream and downstream ends of the interaction region reaching the same point on the detector. This effect is dependent on the radial distance of the detection point from the beam propagation axis. At 9.5 keV X-ray energy and an 86 mm sample-to-detector distance, the $q$ resolution is ~0.06 Å$^{-1}$ or better over the range of detection.

The sizes of the entrance and exit apertures (both 250 µm) were also carefully chosen to not only allow the pump and probe pulses to pass through, but also to allow sufficient flow for proper sample turnover between X-ray pulses. Given that the sample cell is placed inside a vacuum chamber with 2000 L/s of turbomolecular pumping, we can use the approximation that the pressure outside the cell is negligible relative to the pressure inside. Thus, we can calculate the flow rate $q_{cell}$ out of the cell according to

$$q_{cell} = P_{cell} \cdot (A_{entrance} + A_{exit}) \cdot \sqrt{\frac{k_B T}{2\pi m}} \quad (XX.10)$$

where $P_{cell}$ is the pressure inside the cell, $A_{entrance}$ and $A_{exit}$ are the areas of the respective apertures, $k_B$ is the Boltzmann constant, $T$ is the temperature, and $m$ is the molecular mass[42]. Using, for example, 7 Torr of N-methyl morpholine at 22°C, $q_{cell}$ = 0.0427 Torr·L/sec, which means that the cell (with a volume of ~0.4 cm$^3$) will turn over every ~65 milliseconds, or about every eighth X-ray pulse.

In the experiment, the interaction region is a very small portion of the total cell volume. So, it may be more informative to consider the motion of individual molecules as opposed to the collective flow of the ensemble. Thus, we also consider the mean distance $\langle x \rangle$ traveled between X-ray shots,

$$\langle x \rangle = \sqrt{\frac{2\lambda v}{3f}} \quad (XX.11)$$

where $\lambda$ is the mean free path, $v$ is the average thermal particle velocity, and $f$ is the repetition rate of the experiment[42]. Using 7 Torr of N-methyl morpholine at 22°C with the LCLS operating at 120 Hz, we find $\langle x \rangle \approx 825$ µm. Given that the diameter of the X-ray spot is only 30 µm FWHM, the probability of scattering off of the same molecule with multiple X-ray pulses is negligible.

**XX.3.2 Calibration of the Detector Geometry**

In order to properly calibrate the measured absolute scattering signals, it is necessary to consider the physical geometry of the detector relative to the interaction region. Scattering patterns calculated from the Independent Atom Model (IAM) represent the scattering per unit area as a function of $q$ at a fixed distance $R$ between the scattering medium and the point of detection. The IAM allows molecular scattering patterns to be predicted using the atomic form factors, the X-ray wavelength, and the interatomic distances, according to Eq. XX.8.

In the experiments a planar detector is positioned perpendicular to the primary X-ray beam axis. Thus, the distance $R$ depends on the position on the detector, and geometric correction factors must be applied for direct comparison to the IAM. The measured intensity is divided by a $\cos(2\theta)^2$ factor to correct for the $R$ dependence as a function of $2\theta$. The measured intensity is also divided by an additional $\cos(2\theta)$ factor to normalize for the effective area of pixels at different points along the detector. Combined, the measured scattered intensity is divided by $\cos(2\theta)^3$ to compare with calculated IAM patterns[41].

Once the proper correction factors are applied, the measured scattering signal can be compared to the signal predicted from Eq. XX.8 to determine the precise orientation of the detector relative to the interaction region. The detector used is a planar 2.3-megapixel Cornell-SLAC Pixel Array Detector (CSPAD)[43] with a known internal pixel geometry. A least-squares optimization is performed between the IAM image generated from a calculated optimized molecular geometry and the experimentally measured ground-state scattering pattern. The optimization outputs five geometrical parameters. The parameters $x_0$, $y_0$, and $z_0$ are the absolute geometrical coordinates of the center of the detector relative to the interaction region, assuming that the detector plane is perpendicular to the X-ray beam. The fitting also optimizes $\phi_0$, the azimuthal angle of the detector relative to the X-ray polarization, and an overall intensity scaling factor $I_0$. By performing this optimization with ground-state scattering patterns, for which optimized geometries can be calculated reliably, we ensure that the measured excited-state patterns are also properly calibrated.

**XX.3.3 Calibration of timing and intensity jitter**

Using a free-electron laser such as LCLS as the X-ray source has many advantages (such as short pulse duration and high photon flux), but also creates distinct challenges. The LCLS has significant shot-to-shot fluctuations in pulse arrival time, pulse intensity, and spatial orientation. In order to achieve a high signal-to-noise ratio one must account for these fluctuations.

In time-resolved experiments, the relative timing of the pump and probe pulses is controlled via a motorized delay stage. In addition, we also monitor the timing jitter via a spectrally encoded cross-correlator that has been described in detail elsewhere[44]. Briefly, a chirped white light continuum is directed through a thin silicon nitride film, and then dispersed onto a CCD camera as a reference spectrum. Then, the chirped white light is crossed with the X-ray pulse on the film. The X-ray pulse changes the index of refraction of the silicon nitride, causing a drop in transmission. The measured spectrum is then subtracted from the reference spectrum, resulting in a sharp decrease in intensity at a given point in the measured frequency spectrum. After calibrating the position of the edge as a function of X-ray pulse arrival time, this so-called '*time tool*' provides a shot-to-shot measure of the timing jitter.

In the experiments employing the scattering cell described in Sect. XX.3.1, the fluctuations in X-ray intensity incident on the sample arise from two sources: the fluctuations in total pulse-to-pulse X-ray intensity; and the spatial pointing instability of the X-ray beam, which affects transmission through the blocking and entrance apertures. In order to simultaneously correct for both effects, the transmitted X-ray intensity through the sample is monitored with a photodiode downstream of the CSPAD. The single-shot X-ray scattering patterns are then corrected for the photodiode value prior to averaging. It should be noted that this method assumes that the transmitted X-ray intensity is independent of the molecular dynamics. This is a reasonable approximation, as the total probability of an X-ray photon being scattered is ~$10^{-13}$ for dilute molecular vapors. The total transmitted intensity is therefore a very good representation of the total incident intensity and can be used for calibration.

## XX.4 New Observables for Chemical Dynamics and Reactions

Probing molecular dynamics with scattering techniques provides unique views that are often complementary to spectroscopic methods. Consequently, time-resolved scattering can yield new insights about molecular dynamics and kinetics that might be difficult or even impossible to obtain with optical pump-probe spectroscopy. Even though very few time-resolved X-ray scattering experiments have been performed to date, they already have resulted in very important results, some of which are described in the following.

### XX.4.1 Chemical Reaction Dynamics

The first time-resolved X-ray study of chemical reaction dynamics was published in 2015 and explored the chemical ring-opening reaction dynamics of gas phase 1,3-cyclohexadiene (CHD).[40] Excitation at 266 nm places the molecules on the surface of the 1B electronic state, Fig. XX.4.[45] A rapid structural evolution via two conical intersections (CI) transforms the molecule to the open 1,3-hexatriene (HT) form.

The ring-opening reaction of CHD is a good test system for structural dynamics studies because the reaction is perceived to be ballistic[46]: optical excitation creates a well-defined wavepacket that propagates rapidly through the electronic surfaces. Since all molecules in the gas phase sample undergo the same dynamics, the scattering patterns of the molecular ensemble can be equated to the structural dynamics of individual molecules.

The time-dependent scattering pattern is shown in Fig. XX.5. The time resolution of the experiment, as determined from the analysis of the scattering data, was about 80 fs. To determine the time evolution of the molecular structure, trajectory calculations using the multiconfigurational Ehrenfest method were performed with potential energies and nonadiabatic couplings obtained on-the-fly from SA3-CAS(6,4)-SCF/cc-pVDZ *ab initio* electronic structure calculations. Three electronic states are included: the ground state, the optically accessed 1B state, and the 2A state that is implicated in the Woodward-Hoffman mechanism of the electrocyclic reaction. Each trajectory constitutes a molecular reaction path with a complete but distinct set of interatomic distances.

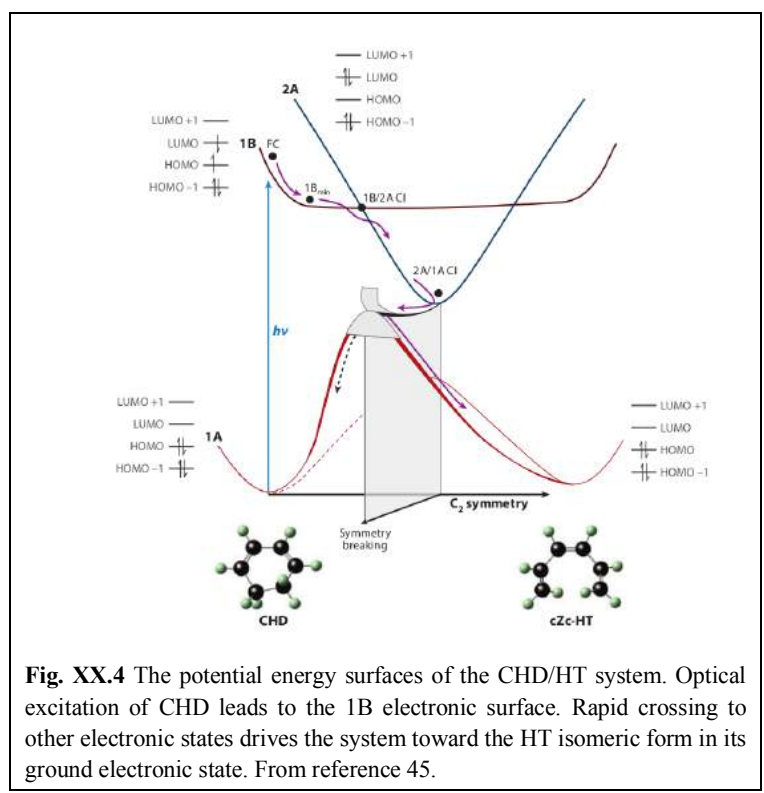

**Fig. XX.4** The potential energy surfaces of the CHD/HT system. Optical excitation of CHD leads to the 1B electronic surface. Rapid crossing to other electronic states drives the system toward the HT isomeric form in its ground electronic state. From reference 45.

To determine the combination of trajectories that best describes the chemical reaction, the experimental scattering patterns were compared to the scattering patterns calculated from the computed trajectories. Optimization converged on a small number of trajectories with four of them having a combined weight of approximately 80%. These trajectories therefore suffice to represent the experimentally observed data, and their graphical representation resulted in 'molecular movies' that show the time-evolving molecular structures. Interestingly, recent simulations by Polyak *et al.*[47] predict a branching ratio between the HT and CHD products closely aligned with that implied by the experiments.

Some of the complexity inherent in the analysis of structures of polyatomic molecules can be avoided by investigating the structural dynamics of diatomic molecules. Glownia *et al*. measured the time-dependent X-ray scattering signals of diatomic $I_2$, Fig. XX.6[48]. Excitation at 520 nm lifts the system to the B state, from where crossing to the B' level can lead to molecular dissociation. The signatures of the dissociation are clearly visible in the plot of the experimental excited state charge distribution *vs*. time, although extracting the charge distribution in real space remains challenging [49],[50],[51]. Dephasing of the vibrational wavepacket leads to a decline in the time-dependent oscillatory signal. The linear polarization of the laser pulse causes an anisotropy of the scattering signal that is readily observed in the experiment and which can be separated from the isotropic component by fitting to a Legendre polynomial.

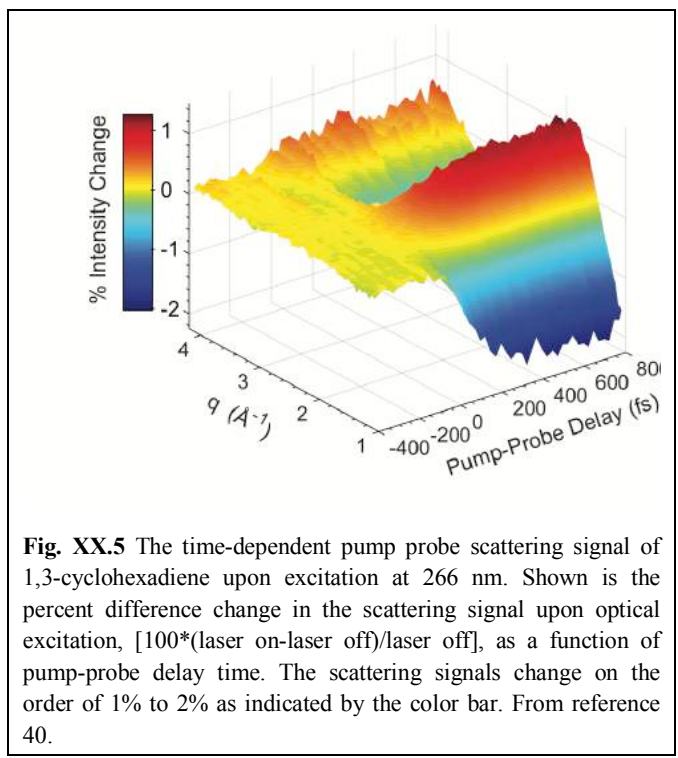

**Fig. XX.5** The time-dependent pump probe scattering signal of 1,3-cyclohexadiene upon excitation at 266 nm. Shown is the percent difference change in the scattering signal upon optical excitation, [100*(laser on-laser off)/laser off], as a function of pump-probe delay time. The scattering signals change on the order of 1% to 2% as indicated by the color bar. From reference 40.

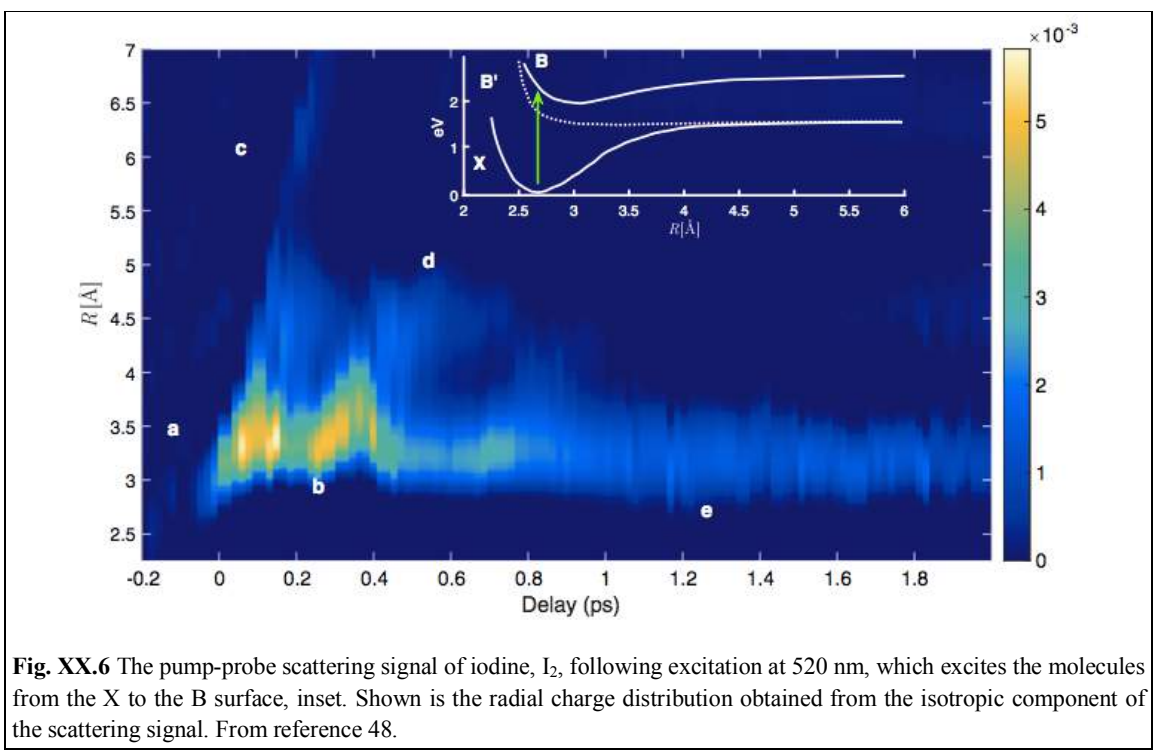

**Fig. XX.6** The pump-probe scattering signal of iodine, $I_2$, following excitation at 520 nm, which excites the molecules from the X to the B surface, inset. Shown is the radial charge distribution obtained from the isotropic component of the scattering signal. From reference 48.

**XX.4.2 Excited State Structures**

Further development of the time-resolved X-ray scattering technique, both experimentally and theoretically, has more recently enabled the determination of polyatomic molecular structures in electronically excited states. In a 2019 study, the time-dependent molecular structure of N-methyl morpholine (NMM) was measured following excitation to a molecular Rydberg state[52], revealing the dephasing of coherent molecular vibrations on a picosecond time scale.

Previous studies of NMM showed that the molecule is promoted to the $3p_z$ molecular Rydberg state following excitation at 200 nm[53,54], which launches a coherent structural oscillation. This oscillation was seen to persist even following internal conversion to the 3s state on a sub-picosecond time scale[53].

Since the structural motions in the excited state are coherent, the distribution of molecular structures in the excited-state ensemble varies approximately normally[52] around a single representative structure. It is this feature that allowed for the development of a novel analysis of time-resolved X-ray scattering patterns that yields complete, time-dependent molecular structures with unprecedented precision.

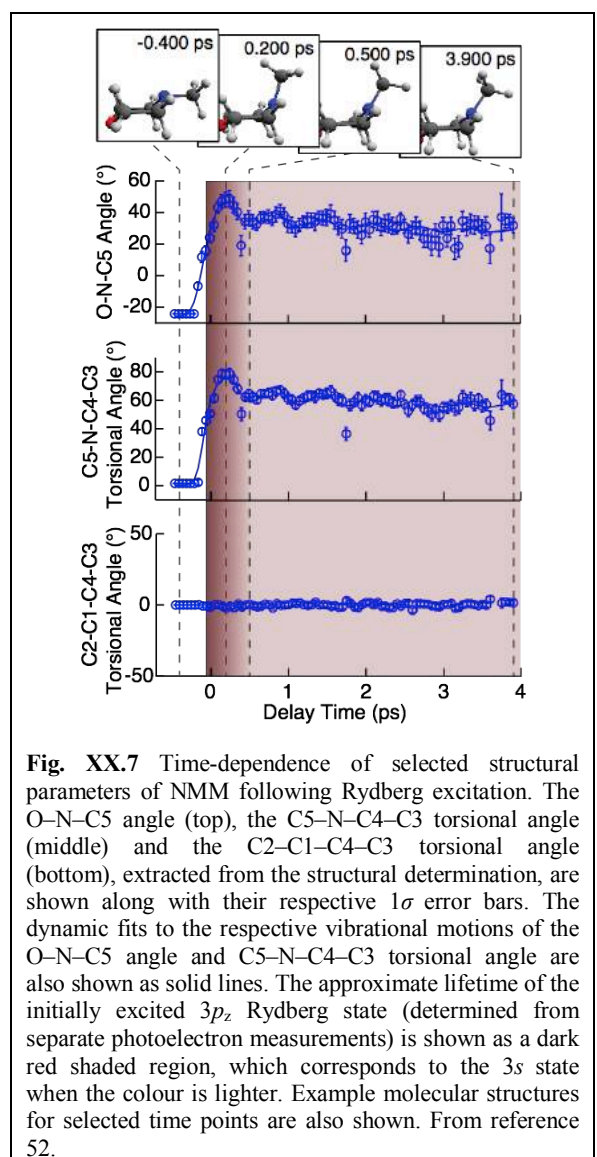

**Fig. XX.7** Time-dependence of selected structural parameters of NMM following Rydberg excitation. The O–N–C5 angle (top), the C5–N–C4–C3 torsional angle (middle) and the C2–C1–C4–C3 torsional angle (bottom), extracted from the structural determination, are shown along with their respective $1\sigma$ error bars. The dynamic fits to the respective vibrational motions of the O–N–C5 angle and C5–N–C4–C3 torsional angle are also shown as solid lines. The approximate lifetime of the initially excited $3p_z$ Rydberg state (determined from separate photoelectron measurements) is shown as a dark red shaded region, which corresponds to the 3s state when the colour is lighter. Example molecular structures for selected time points are also shown. From reference 52.

The analysis begins with calculation of a large set of surface-hopping trajectories propagated on the ground and excited electronic states, with the appropriate amount of excess kinetic energy included. While each trajectory is unique, all trajectories sample physically reasonable and energetically accessible regions of the structural phase space. From the trajectories, approximately one million individual molecular geometries were extracted, from which a corresponding set of scattering patterns were calculated.

Using this pool of calculated scattering patterns, the least-squares fitting errors of each pattern compared to the experimentally measured signal are calculated at each time point. Given that the least-squares fitting error was observed to vary approximately normally as a function of each individual structural parameter, the best-fit values of all structural parameters are obtained independently at each time point. This results in a full set of molecular structure parameters (*i.e.* the molecular structure) at each time point, which are then viewed in succession to reveal a detailed picture of the time-dependent molecular motions. Selected representative structural parameters of NMM as a function of time are shown in Figure XX.7.

**XX.4.3 Correlated Structures far from Equilibrium**

While scattering experiments measure the electron and charge density distributions in molecules, the concept of '*molecular structure*' itself merits careful thought. In reacting systems, the chemical dynamics oftentimes starts out as a wavepacket with a well-defined maximum that resembles a classical system. But during a reaction there is a large amount of energy in play, which typically is distributed into a hot bath as the molecule approaches to a thermal state. The effect of this energy redistribution is readily seen in the experiments on $I_2$ and NMM discussed above, Figures XX.6 and XX.7.

Hot molecular systems may have large-amplitude vibrations that can, depending on the structural rigidity of the molecule, lead to geometries that are far from equilibrium. To describe their scattering signals, distance shifts, anharmonicities and structural correlations need to be considered.[55] The anharmonicity of the interatomic or normal mode potentials causes average bond distances to be longer than the equilibrium distance. A vibrationally hot molecule therefore has a different scattering pattern than a cold molecule, Figure XX.8.

Depending on the molecular geometry, interatomic distances may shrink or increase with vibrational excitation. The inset in Fig. XX.8 illustrates the effect on the example of a triatomic system, ABC. Vibrational excitation of the bending vibration causes the average distance between atoms A and C to be less than the sum of the A-B and B-C distances. The effect on the scattering pattern can be substantial, as Fig. XX.8 shows in the example of the CHD molecule.

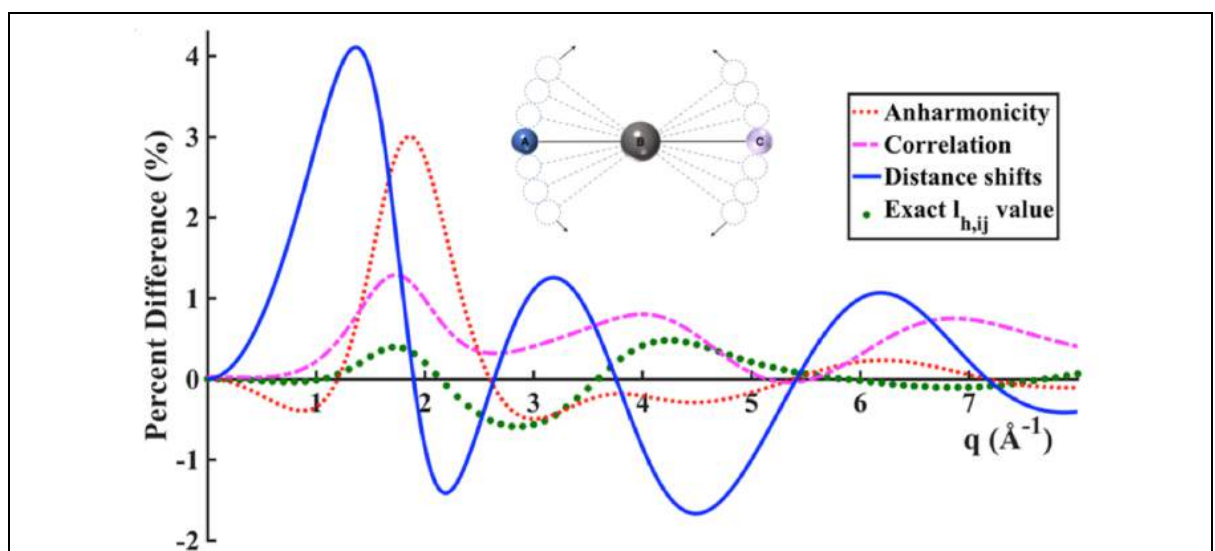

**Fig. XX.8** Contributions to the percent difference scattering signals of the CHD model system, after excitation with 6 eV photons and subsequent relaxation into thermal (2870 K) vibrations. Shown are the contributions arising from the distance shifts, anharmonicity, correlations between atom-atom pair distances and exact vibrational amplitudes $l_{h,ij}$. From reference 55.

Depending on the scattering vector, it may be much larger than the more intuitive changes arising from complicated structure distributions associated with complex potential energy surfaces far from the equilibrium structure ('exact vibrational amplitudes $l_{h,ij}$').

The Debye formula, given in Eq. XX.8, expresses the scattering signal from interferences due to the distance between pairs of atoms. In an $N_{at}$-atom molecule, there are ½$N_{at}(N_{at}-1)$ such interatomic distances yet there are only $3N_{at}-6$ normal mode coordinates. For any molecule with more than 4 atoms, there are therefore fewer normal modes than interatomic distances. Consequently, there must be redundant parameters, implying that correlations between interatomic distances need to be considered for vibrationally very hot molecules. Figure XX.8 illustrates that the effect can be considerable, although not quite as large as the contributions from the distance shifts and the anharmonicity. A complete analysis, considering all the described effects, can simulate scattering patterns that are in excellent agreement with experimentally measured patterns.[55] Importantly, the analysis of the NMM excited state structure described above inherently takes structural correlations, as well as the other effects, into consideration.

## XX.4.4 Chemical Kinetics

X-ray scattering experiments have recently also proven to be powerful tools for investigating excited-state intermediates during kinetic processes. In contrast to dynamic experiments, which probe the molecular structure of a reacting molecule on the timescale of the nuclear motions, kinetic experiments seek to measure the relative populations of reactant and product species as a function of time. In the case of long-lived intermediates, the measured scattering signals include the characteristic pattern of the intermediates, and so their population can also be monitored.

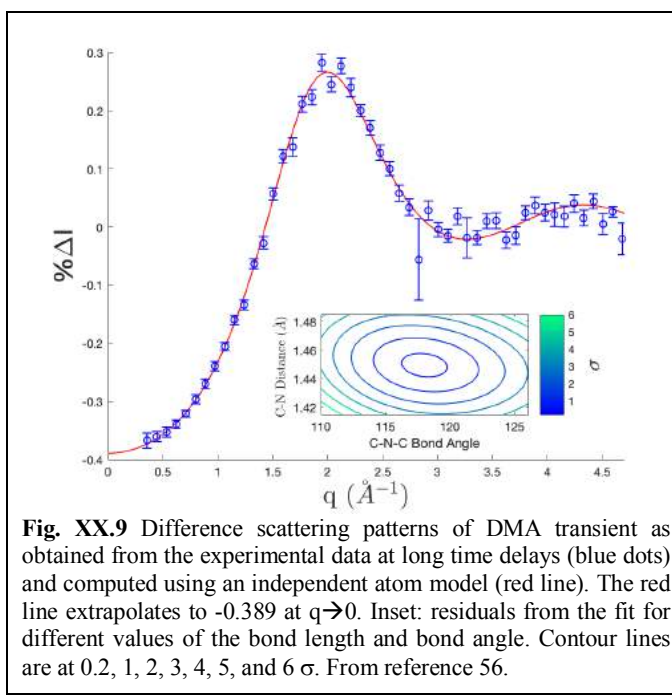

**Fig. XX.9** Difference scattering patterns of DMA transient as obtained from the experimental data at long time delays (blue dots) and computed using an independent atom model (red line). The red line extrapolates to -0.389 at q→0. Inset: residuals from the fit for different values of the bond length and bond angle. Contour lines are at 0.2, 1, 2, 3, 4, 5, and 6 σ. From reference 56.

Recently, X-ray scattering was used to investigate the photodissociation of a prototypical tertiary amine, trimethylamine (TMA)[56]. Previous photoionization and molecular beam studies showed that when TMA is optically excited near 200 nm, the $3p_z$ molecular Rydberg state is prepared, which quickly undergoes internal conversion to $3p_x$ and $3p_y$, which subsequently decay into the 3s Rydberg state on a picosecond time scale[57,58]. In addition, Forde *et. al.* measured the product distributions following excitation at 193 nm, which identified two significant photodissociation channels: the dominant one involving a stepwise reaction to form N-methylmethanimine (NMMA), and a secondary channel forming dimethylamine (DMA) and methyl radicals[59,60]. Left unknown, however, were the kinetic timescales for product formation and the molecular geometries of the product species.

In the study by Ruddock *et. al.*[56], TMA was excited by a 200 nm optical pulse and probed via hard X-ray scattering. The scattering patterns revealed two distinct photodissociation pathways: a minor fast dissociation pathway from the 3p state, and a dominant slower dissociation which proceeds via internal conversion to the 3s state. Both pathways were found to yield dimethylamine and methyl radicals, with no direct evidence for the formation of NMMA. In addition, the scattering signals at long delay times showed a clear signature of the DMA radical (see Fig. XX.9), which allowed for extraction of a detailed molecular structure of this transient.

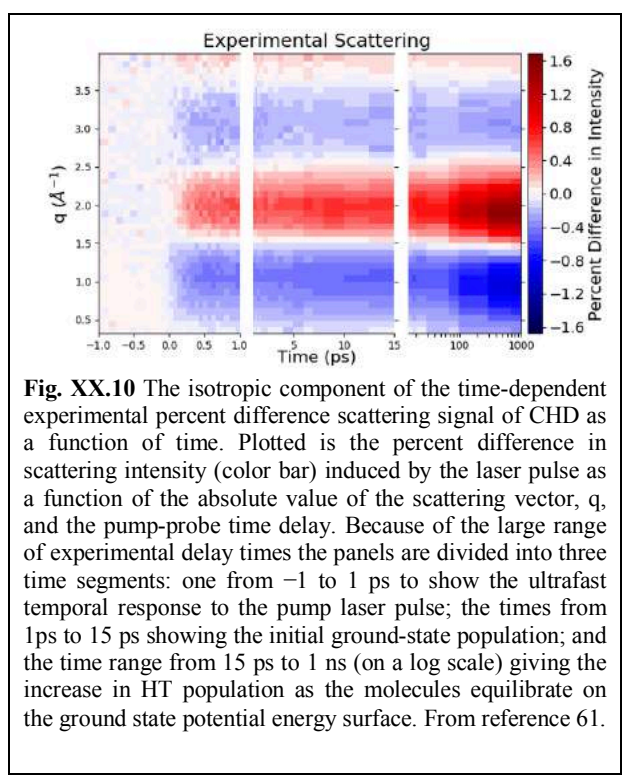

**Fig. XX.10** The isotropic component of the time-dependent experimental percent difference scattering signal of CHD as a function of time. Plotted is the percent difference in scattering intensity (color bar) induced by the laser pulse as a function of the absolute value of the scattering vector, q, and the pump-probe time delay. Because of the large range of experimental delay times the panels are divided into three time segments: one from −1 to 1 ps to show the ultrafast temporal response to the pump laser pulse; the times from 1ps to 15 ps showing the initial ground-state population; and the time range from 15 ps to 1 ns (on a log scale) giving the increase in HT population as the molecules equilibrate on the ground state potential energy surface. From reference 61.

More recently, the same technique was used to investigate the kinetics of cyclohexadiene (CHD) following excitation at 200 nm to the 3p molecular Rydberg state[61]. In addition to the ultrafast experiments studying CHD excited near 267 nm as previously discussed, the molecular response of CHD to deep UV excitation near 200 nm has also been the subject of several investigations. Ultrafast photoelectron experiments by Bühler *et. al.*[62] revealed that the initially excited Rydberg state quickly decays, but the experiments showed no evidence of ring opening. In contrast, Sekikawa and coworkers[63,64] used two-photon excitation with 400 nm pulses to investigate the same region of the absorption spectrum, and suggested that the decay of the initially excited state initiated ring opening on a sub-picosecond time scale. Given the lack of consensus on the molecular structure following decay of the 3p state, a more definitive determination of the time-evolving molecular structure was deemed necessary.

In 2019, Ruddock *et. al.*[61] used time-resolved X-ray scattering to probe the molecular structure of CHD following excitation with 200 nm optical pulses. The difference scattering signals were measured from the early femtosecond time scale out to 1 ns (see Fig. XX.10), providing a full picture of the reaction progress.

Analysis of the measured scattering patterns showed that the initially excited 3p$_x$ and 3p$_y$ Rydberg states quickly find their way to the electronic ground state, where initially 76% of the molecules are in the closed-ring CHD form, with the remainder having undergone rapid ring-opening to form the hexatriene (HT) product. Then, the hot ground-state CHD undergoes a thermal reaction to form the HT product on a 174 ps time scale[61] (see Fig. XX.11). In addition to determining the kinetic parameters for this reaction, further analysis of the product scattering pattern revealed a thermal distribution of rotameric forms in the HT product. In addition to being a powerful tool for studying ultrafast reactions, time-resolved X-ray scattering has thus proven to be a valuable technique for determining reaction rates, revealing multiple reaction pathways, and precisely measuring product molecular structures in kinetic reactions.

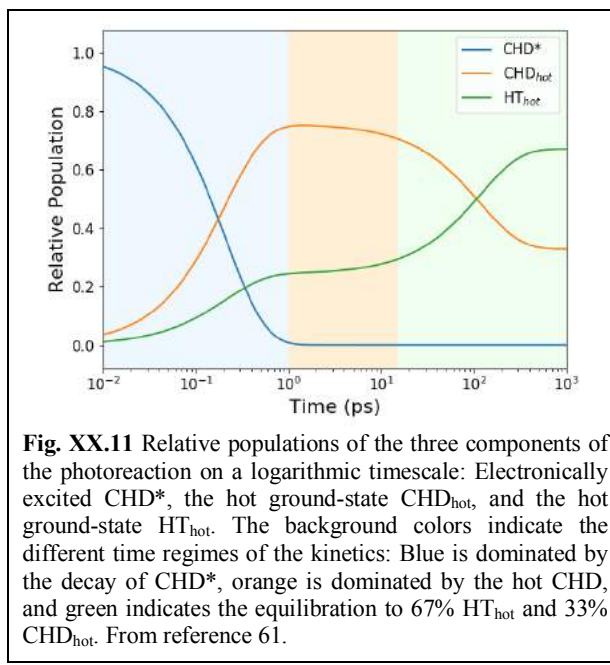

**Fig. XX.11** Relative populations of the three components of the photoreaction on a logarithmic timescale: Electronically excited CHD*, the hot ground-state CHD$_{hot}$, and the hot ground-state HT$_{hot}$. The background colors indicate the different time regimes of the kinetics: Blue is dominated by the decay of CHD*, orange is dominated by the hot CHD, and green indicates the equilibration to 67% HT$_{hot}$ and 33% CHD$_{hot}$. From reference 61.

### XX.4.5 Counting Electrons

For small scattering angles, the scattered amplitudes from different electrons in a molecule add coherently. As the scattering vector approaches zero, the X-ray scattering signal of a molecule with $N$ electrons is proportional to $N^2$, a result of the coherent addition of amplitudes[56],

$$S(q \to 0) = N_e^2. \qquad (XX.12)$$

Free molecules in the gas phase are at random separations, so that scattered amplitudes from different molecules add up incoherently. When absorption of a photon fragments a molecule into components with electron counts $N_\alpha$, with $N_e = \sum_\alpha N_\alpha$, the scattering signal close to $q = 0$ is proportional to the sum of the fragment electron counts squared,

$$S(q \to 0) = \sum_\alpha N_\alpha^2 . \qquad (XX.13)$$

The sum of the squares is always smaller than the square of the sum, and consequently any dissociation is associated with a reduction in the X-ray scattering signal for very small scattering vectors. Experimentally this was observed on the dissociation of trimethyl amine (TMA) to methyl radicals and dimethyl amine (DMA). Figure XX.12 shows the scattering signals at various delay times. In the sub-picosecond regime, the molecules are bonded, as witnessed by the convergence of the percent difference scattering signal toward zero with small $q$, even though the optical excitation at 200 nm changes the scattering signals at larger $q$. Since TMA has 34 electrons, the scattering signal extrapolated to $q\rightarrow 0$ is proportional to $34^2 = 1156$ for the undissociated molecules.

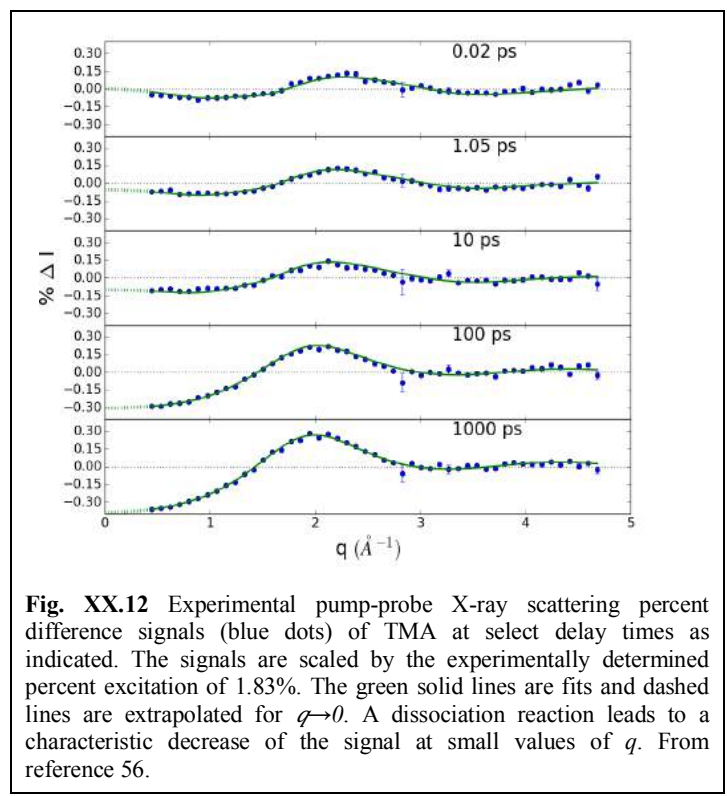

**Fig. XX.12** Experimental pump-probe X-ray scattering percent difference signals (blue dots) of TMA at select delay times as indicated. The signals are scaled by the experimentally determined percent excitation of 1.83%. The green solid lines are fits and dashed lines are extrapolated for $q\rightarrow 0$. A dissociation reaction leads to a characteristic decrease of the signal at small values of $q$. From reference 56.

As the fragmentation proceeds with increasing time delays the percent difference scattering signal approaches a lower value. The methyl radical has 9 electrons, and DMA has 25, so that the sum of the scattering signals becomes proportional to $25^2 + 9^2 = 706$. Therefore, once the reaction to DMA and $CH_3$ is complete, the $q\rightarrow 0$ percent difference signal is expected to reduce by $100(706 - 1156)/1156 = -38.9\%$ compared to the TMA molecule. This is exactly what is observed in the experiment, shown in Fig. XX.12.

The quantitative measurement of the scattering signals at very low scattering angles can therefore provide an accurate determination of the electron counts of the fragments involved. This can be very useful for the identification of fragmentation pathways and kinetic reaction schemes. In the case of TMA, it led to the discovery of a fast reaction channel that had previously not been identified.

A persistent question in pump-probe scattering experiments relates to the identity of the initially excited state. For any optical excitation that creates a sizeable population in an excited state, there is the possibility that absorption of additional photons leads to yet higher electronic states. For many experiments with electronic excitation in the UV part of the spectrum, absorption of a second photon would lead to the ionization of the molecule. A molecule with $N_e$ electrons would become a cation with $N_e - 1$ electrons, so that the small-angle scattering signal would be reduced from $N_e^2$ to $(N_e - 1)^2 + 1^2$. For many prototypical chemical reaction dynamics systems, the effect is in the range of several percent, so that a careful measurement of the pump-probe scattering signals can experimentally determine whether a significant fraction of the molecules have absorbed a second photon.

**XX.4.6 Electron Density Distributions**

As previous examples illustrate, scattering can reveal structure and structural changes in molecules. In the context of X-ray scattering, this is because a significant portion of the signal tracks the positions of nuclei, predominantly via scattering from the tightly bound core electrons. However, scattering occurs from *all* electrons and the valence electrons also contribute. These can be significantly distorted by chemical bonding. The discrepancy between a nucleus-centered electron distribution and actual observations was first observed in high-resolution X-ray crystallography, in particular for light elements such as hydrogen. In those situations, an expansion of the electron density using a multipole model was found to provide a more accurate description of the experimentally observed scattering patterns and the associated electron density[65].

Ultrafast dynamics involves excited electronic states, which have electron distributions distinct from the ground electronic state. The initial change in electronic structure due to vertical excitation by the pump pulse provides one of the cleanest examples of this. An early theoretical study by Ben-Nun *et al.* investigated the change in elastic X-ray scattering due to excitation of a lithium atom from the $^2$S electronic ground state to the first excited $^2$P state[66]. Comparing the calculated signal for the ground state versus the excited state, shown in Fig. XX.13, the effect of the redistribution of electron density upon excitation is seen clearly in the scattering pattern. More recently, Northey *et al.* considered the changes in the elastic scattering upon vertical excitation in molecular systems[9]. In Fig. XX.14 the calculated elastic scattering from the excited $S_1$ state in the molecule 1,3-cyclohexadiene is shown. The magnitude of the changes in the scattering pattern are comparatively small, as seen in the lower panel which shows the difference in scattering between the $S_1$ excited and the $S_0$ ground state. This relates to the comparatively minor changes in electron density, in first approximation on the order of one electron moving between *e.g.* the highest occupied and the lowest unoccupied molecular orbital. However, although the changes are small in absolute terms, as a relative percent difference change they can be significant[9,10], with the change at specific pixels in the current example reaching 40% for an aligned molecule.

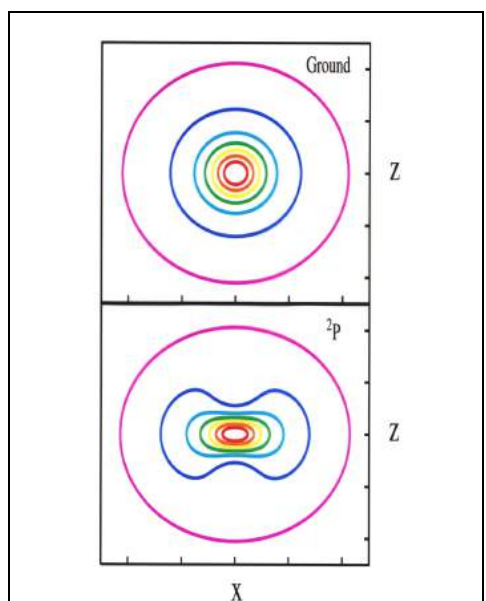

**Fig. XX.13** Calculated elastic X-ray scattering for Li atoms in the ground state (upper panel) and when excited by a linearly polarized photon to the $^2$P state (lower panel). The change in the electron density distribution upon excitation leads to a change in the X-ray scattering intensity. From reference 66.

Experimental detection of changes in electronic structure via X-ray scattering requires high-quality data. Early experiments could quite adequately be interpreted using the independent atom model (IAM), which only accounts for the electron density of a molecule in an approximate manner and cannot describe excited electronic states, as discussed in Sect. XX.2. The recent experiment by Stankus *et al.*[52] is indicative of the advances that have been made since. In that experiment excited state structural dynamics in the molecule NMM was measured (see Sect. XX.4.2). Interpretation of the data required that the IAM was augmented by a correction from electronic structure theory that accounted for the change in electron density due to the molecular dynamics evolving on diffuse excited $3p_z$ and $3s$ Rydberg states. However, the measurement was insufficient to positively identify the electronic state along the lines of previous theoretical predictions in atoms.[67,68] As Fig. XX.15 shows, the correction due to electron density redistribution is small compared to the changes in the signal resulting from changes in the molecular geometry, but is not negligible. Going forward, it is likely to become increasingly important to use accurate models that account for the electronic structure when interpreting experimental data. This work is well underway. Codes that calculate elastic scattering from ground and excited states have been developed[9,10], and can handle larger molecules [69]. The codes can also compute scattering due to specific inelastic transitions[13] as well as total scattering[3]. To be practical for the interpretation of large volumes of experimental data, these methods have to be computationally efficient[4,5].

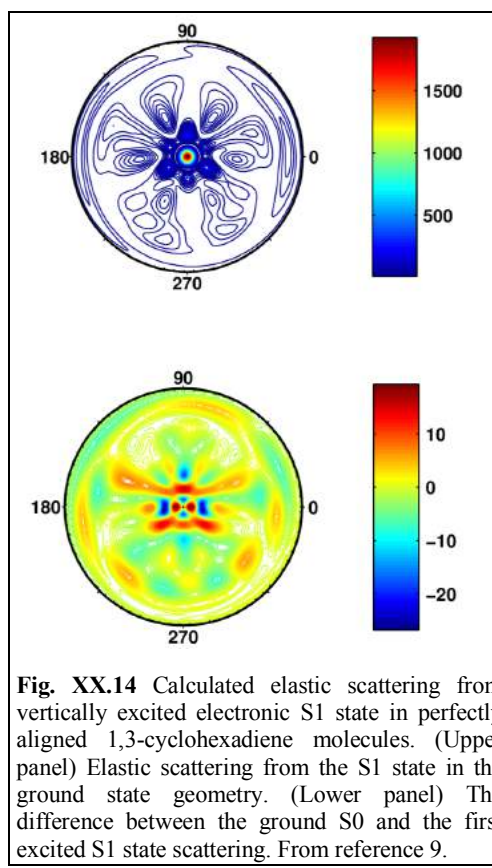

**Fig. XX.14** Calculated elastic scattering from vertically excited electronic S1 state in perfectly aligned 1,3-cyclohexadiene molecules. (Upper panel) Elastic scattering from the S1 state in the ground state geometry. (Lower panel) The difference between the ground S0 and the first excited S1 state scattering. From reference 9.

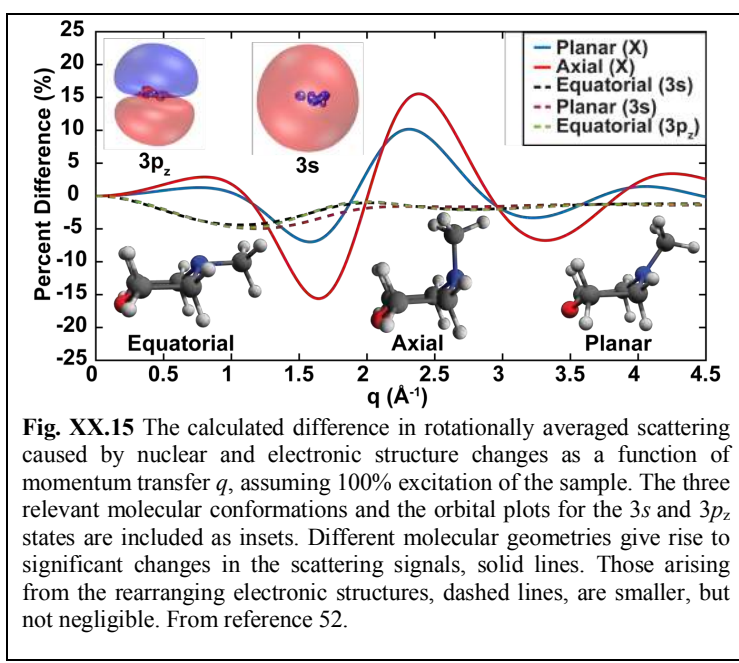

**Fig. XX.15** The calculated difference in rotationally averaged scattering caused by nuclear and electronic structure changes as a function of momentum transfer *q*, assuming 100% excitation of the sample. The three relevant molecular conformations and the orbital plots for the $3s$ and $3p_z$ states are included as insets. Different molecular geometries give rise to significant changes in the scattering signals, solid lines. Those arising from the rearranging electronic structures, dashed lines, are smaller, but not negligible. From reference 52.

Looking further ahead, it does appear likely that ultrafast X-ray scattering experiments will eventually succeed to track not only the initial change in electron density due to excitation by the optical pump pulse, which would be an important result in itself, but also subsequent transitions between electronic states due to nonadiabatic or spin-orbit coupling as the molecule relaxes towards the product states. Such experiments would close the gap between ultrafast structural dynamics and existing ultrafast spectroscopies. A particularly attractive aspect of such experiments would be the ability to track structural and electronic changes in a single experiment. However, challenges remain. Apart from requiring experimental signals of very high quality and methods to predict the scattering for various electronic states, one would have to tackle the problem of separating contributions to the scattering signal from changes in electronic structure, changes in population, and molecular geometry. This is a nontrivial task. The most likely route forward in the immediate future is therefore experiments whose interpretation relies on a combination of scattering and spectroscopic data[70,71]. There is also interesting overlap with quantum crystallography[72,73], which grapples with similar challenges in the context of high-quality static X-ray diffraction data from ground state molecules.

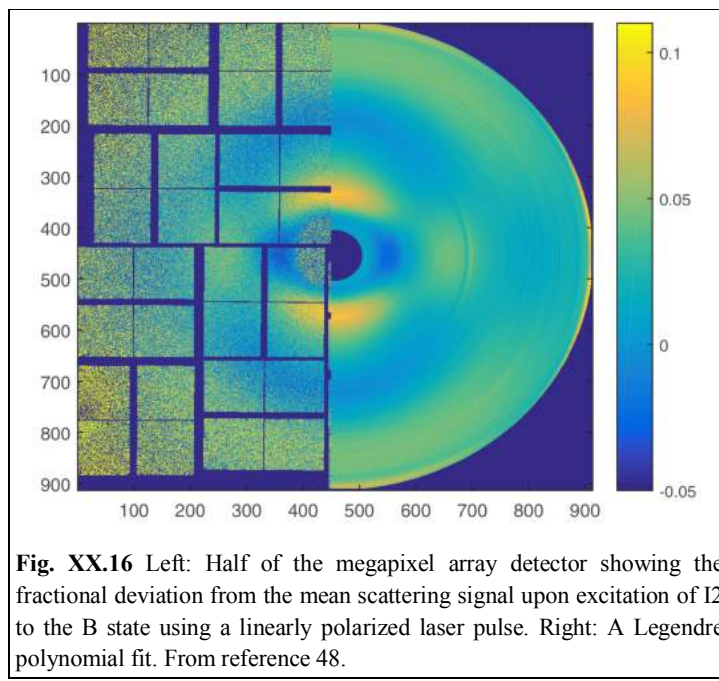

**Fig. XX.16** Left: Half of the megapixel array detector showing the fractional deviation from the mean scattering signal upon excitation of I2 to the B state using a linearly polarized laser pulse. Right: A Legendre polynomial fit. From reference 48.

### XX.4.7 Transition Dipoles

Most optical pump, X-ray probe experiments utilize linearly polarized optical laser radiation. The interaction of the linearly polarized light with the randomly oriented molecules in the gas phase selects out those molecules whose transition dipole moments align with the laser polarization. This induces an anisotropy that is readily detected in the scattering experiments.

Figure XX.16 illustrates this in the example of the diatomic $I_2$ system[48]. Excitation with a linearly polarized laser pulse at 520 nm leads to a mix of optically excited and ground state molecules. The distribution of both is anisotropic, as the laser excitation has removed molecules from the ground state and created the excited state population. Since the scattering signatures of ground and excited state molecules are different, the overall scattering signal reflects the anisotropy of the molecular ensemble.

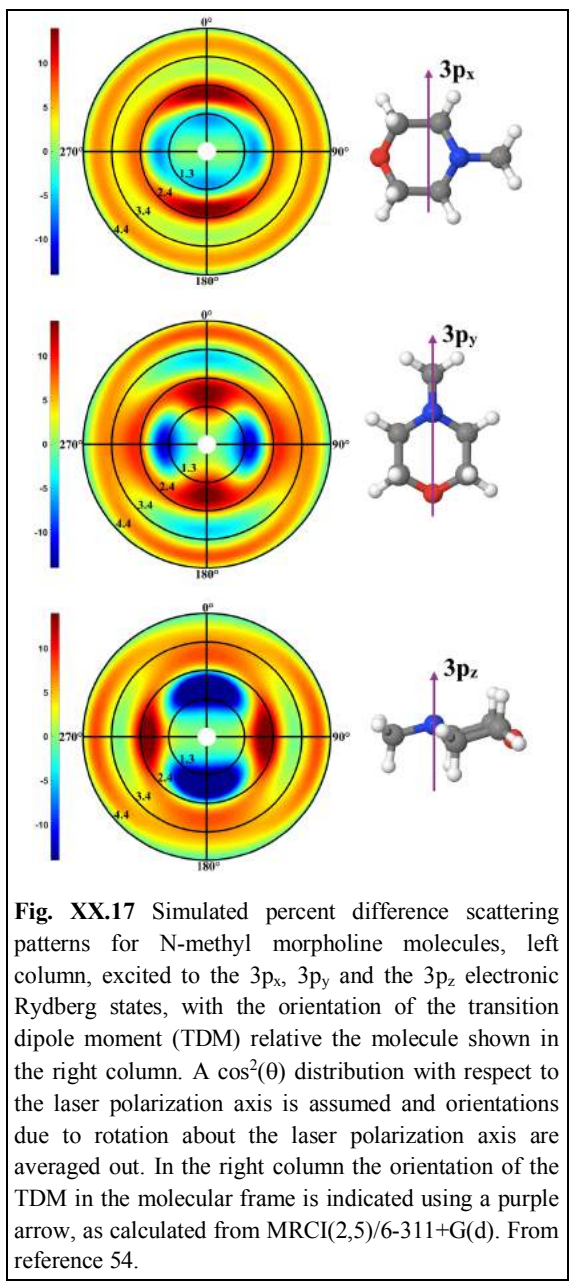

**Fig. XX.17** Simulated percent difference scattering patterns for N-methyl morpholine molecules, left column, excited to the $3p_x$, $3p_y$ and the $3p_z$ electronic Rydberg states, with the orientation of the transition dipole moment (TDM) relative the molecule shown in the right column. A $\cos^2(\theta)$ distribution with respect to the laser polarization axis is assumed and orientations due to rotation about the laser polarization axis are averaged out. In the right column the orientation of the TDM in the molecular frame is indicated using a purple arrow, as calculated from MRCI(2,5)/6-311+G(d). From reference 54.

The anisotropy of the scattering signals turns out to be a very useful tool to determine the excited state prepared in the optical excitation. The alignment of the optical transition dipole moment, as reflected in the scattering patterns, provides important information on the symmetries of the involved states, and can therefore help identify the initially prepared state that is the subject to the observed chemical dynamics.

Figure XX.17 illustrates the concept in the NMM molecule[54]. Excitation in the 200 nm region might excite any of the 3p Rydberg states, or a mixture thereof. But since the sublevels of 3p align differently with respect to the frame of the molecule, the resulting pump-probe scattering patterns would be different. A comparison to the experiment, Fig. XX.18, unambiguously determines that the optical excitation is predominantly to the $3p_z$ level, which therefore is the starting point of the ensuing chemical dynamics.

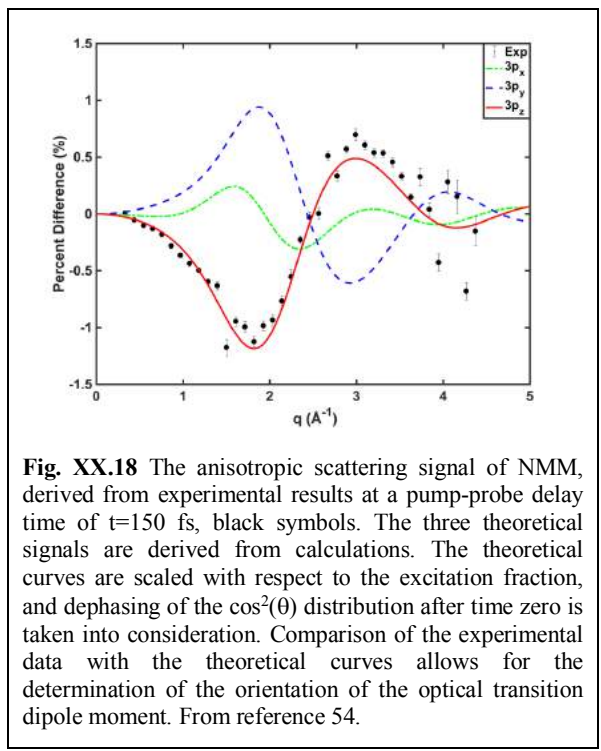

**Fig. XX.18** The anisotropic scattering signal of NMM, derived from experimental results at a pump-probe delay time of t=150 fs, black symbols. The three theoretical signals are derived from calculations. The theoretical curves are scaled with respect to the excitation fraction, and dephasing of the $\cos^2(\theta)$ distribution after time zero is taken into consideration. Comparison of the experimental data with the theoretical curves allows for the determination of the orientation of the optical transition dipole moment. From reference 54.

## XX.5 Outlook

Scattering experiments are beginning to contribute a new perspective on the dynamics of molecular reactions. The tremendous intensity of X-ray free electron lasers, paired with the ability to deliver pulses with durations rivaling those of optical lasers, has made the investigation of chemical dynamics of free molecules possible using time-resolved X-ray scattering. Important insights have already been obtained, as described in this chapter. Without doubt, new XFEL technologies will continue to drive advances that will lead to a deeper understanding of the basic functioning of molecules and their transformations.

Most time-resolved X-ray scattering experiments to date have been performed with X-ray photon energies below 10 keV, limited by the range of the fundamental beam of the LCLS light source. Upcoming upgrades will extend the range to include higher photon energies. This will be an important step, because it will enable the measurement of larger scattering vectors, where the signatures of the detailed shapes of wavepackets are expected to be found.[74] These features occur on small real-space length scales for prototypical reactions such as the ballistic motions of CHD. Consequently, they will appear at higher scattering vectors in the scattering signals, Fig. XX.19. Directly observing the shape and time evolution of dynamical wavepackets would be an important advance in our exploration, and eventual control, of chemical reactions.

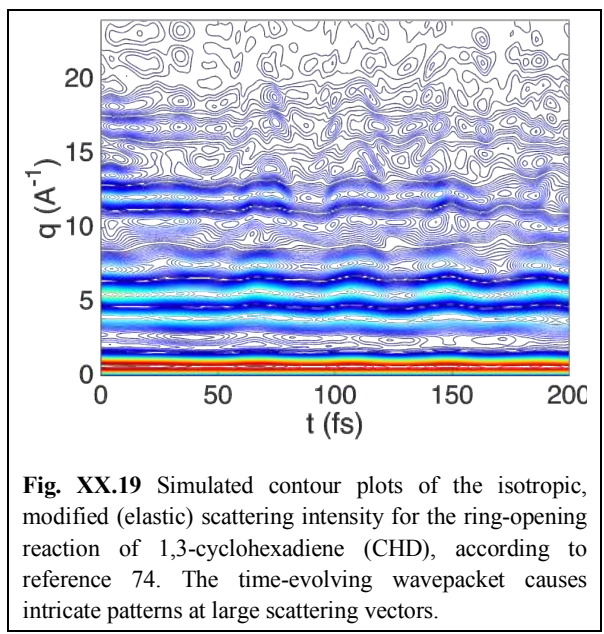

**Fig. XX.19** Simulated contour plots of the isotropic, modified (elastic) scattering intensity for the ring-opening reaction of 1,3-cyclohexadiene (CHD), according to reference 74. The time-evolving wavepacket causes intricate patterns at large scattering vectors.

The manner in which scattering investigations of chemical dynamics complement spectroscopy is an important motivation for further development of the scattering methods. Intriguingly, while ultrafast spectroscopy has been developed for several decades and has become very powerful, it may well be ultrafast scattering that ultimately delivers more insight. As discussed in Sect. XX.4.6, we may eventually be able to assign electronic states during the dynamics based on analysis of the actual electron distribution, and theoretical predictions of coherent effects in scattering indicate that it may be possible to access information on electronic transitions, transient electron dynamics, and nuclear coherences[75,76,77,78]. Furthermore, all time-resolved spectroscopic experiments are limited by the Heisenberg uncertainty relation, which imposes a fundamental limit on measuring the energies of spectroscopic transitions and time simultaneously. A scattering experiment ultimately yields positions, in form of charge or electron density distributions. Since the Heisenberg uncertainty relation imposes no fundamental limit on the simultaneous measurement of positions and time, it may be that, once the technologies are fully developed, scattering experiments may give us the most detailed views of chemical dynamics.